\documentclass[12pt]{article}

\usepackage[dvips]{graphicx, color}
\begin{document}

\title{Investigation of a 90 Degree Spherical Deflecting Analyzer Operated in
an Asymmetrically Charged Configuration}
\author{Chong-Yu Ruan, Scott Nguyen, Manfred Fink\\
   Department of Physics, University of Texas, Austin, Texas 78712, USA}
\date{}
\maketitle
\begin{abstract}
The electron optics of a 90 degree spherical deflecting analyzer (SDA-90) is
investigated with an imaging matrix formalism. As a preanalyzer in the UTA-neutrino 
experiment, high transmission and reasonable energy resolution are the choices of 
optimization. The magnification of the source through the analyzer plays the key
 role in determining the energy resolution.  The imaging matrix approach provides 
graphical information to facilitate such an evaluation. We can demonstrate that in
 case where the analyzer is asymmetrically charged, the rotation of the image 
helps increase both transmission probability and resolution. A telefocus electron 
gun is used to check the numerical result, and to investigate the transverse 
focusing behavior.
\end{abstract}
 
\section{Introduction}

Direct measurements of the neutrino mass face the challenge of interpreting a
convoluted spectrum to very high precision. The negative mass square issue prevails 
\cite{Lobashev94} \cite{Bach93} \cite{Decman91} \cite{Robertson91}
 while the request for higher resolution presses on. The spectrometer function 
plays a key role in deciphering the mystery, but often the convolution of the 
finite source volume is not easy to take into account\cite{Wilkerson93}.
 In the UTA experiment, 
the resolution is set to reach the $10^{-5}$ level while sufficient counts must be 
recorded to suppress statistical uncertainties. A 90 degree spherical 
analyzer (SDA) is designed as a preanalyzer for the UTA neutrino mass
experiment. This analyzer has to provide a $\pm 3.5^{\circ }$ acceptance cone for the
beta particles (electrons) emitted from a cell positioned along the
symmetrical axis. All the emission from within the fudiciary source volume is expected to
 be imaged through a narrow ring slit. This image will serve as the source 
for a high resolution cylindrical mirror analyzer(CMA). The ring slit of the SDA controls the 
flow of tritium gas emanating from the cell and cuts out the low or high energy tail of 
the distribution function for the CMA. The SDA-90 provides high luminosity, narrow 
throughput image and reasonable energy resolution, characterizing a focusing analyzer. 
Using the SDA as a focusing instrument was
first proposed by Aston \cite{Aston19} and later investigated by 
Purcell analytically based on trajectory analysis\cite{Purcell38}. Ashby then 
included relativistic corrections \cite{Ashby58}, but left out fringe effects. 
Kessler et al. formulated a mathematical model for the fringe
 effects, but due to the limitation of their instrument, only first order focusing
 was seen in their experiment\cite{Kessler64}. The second order focusing was included
 in the design of Ross et al. by adding two Herzog lenses to adjust the fringe 
fields\cite{Ross94}. An important feature of that design was that the analyzer was 
asymmetrically charged. In this investigation we verify that we can maintain 
second-order focusing which is not sensitive to the positions of emitters. Regarding 
the spherical aberration as a minor effect, the imaging property of the analyzer has  
been further investigated.  A SDA of very similar
 design by Ross et al. has been built, and our simulation results were checked using a 
telefocus electron gun as the source. 

\section{Theoretical Background}

Theoretical studies of the electron optics of an analyzer is generally based on
trajectory analysis. When an analytic form of a trajectory is available,
it generally can be represented in the form
\begin{eqnarray}
L=L\left( \theta ,n,k\right)
\label{eq:L}
\end{eqnarray}
where $L$ is the projection of the flight path from the source to the image
onto the symmetric axis, $\theta $ is the azimuthal angle of
the incident trajectory, $n$ characterizes the source position, and $k$
accounts for the voltage configuration of the analyzer and the kinetic energy of 
the electrons. Similar to the optical axis in light optics, the principle trajectory is 
defined as the orbit of the electrons which goes along the geometrical central path from
the source through the analyzer. For a point emitter, the deviations from the principle
trajectory due to angular dispersion and energy dispersion can be expressed
through a Taylor expansion
\begin{eqnarray}
\Delta L(\Delta \theta, \Delta E )={\sum_{\mu=1}^{\infty}}\frac{1}{\mu !}\left( 
\frac{\partial ^{\mu }L}{\partial \theta ^{\mu }}\right) _{L_{0}}\left(
\Delta \theta \right) ^{\mu }+{\sum_{\nu=1}^{\infty} }
\frac{1}{\nu !}\left( \frac{\partial ^{\nu }L}{\partial E^{\nu }}\right)
_{L_{0}}\left( \Delta E\right) ^{\nu }+R
\label{eq:Taylor}  
\end{eqnarray}
The first term of Eq.(\ref{eq:Taylor}) characterizes the spherical aberration, while 
the second term characterizes the energy dispersion in the image plane.  
The `mixed' term $R$ is generally not important if the spectra to be analyzed is
 not continuous over a large range.  The energy dispersion is defined as
\begin{eqnarray}
D=E_{0}\left( \partial L/\partial E\right)  
\label{eq:dispersion}
\end{eqnarray}
Often the source can not be considered as a point. For a finite source ( size $l$ )
 the resolution of the analyzer can be modeled as
\begin{eqnarray}
R=\frac{E}{\Delta E}=\left[ \frac{M_l l}{D}+\frac{\Delta L\left( \Delta \theta \right) }{D}\right]^{-1}, 
\label{eq:resoln1}
\end{eqnarray}
and the transmission density is given by
\begin{eqnarray}
  \label{eq:trans}
  T=\frac{N' \Omega}{8\pi^{2} \left( M_l l+\Delta L\right) R}
\end{eqnarray}
where $N'$ is the number of total electrons emitted per sec, $\Omega$ is the solid angle of
the acceptance cone, $M_l$ is the lateral magnification, and $R$ is the radius of the ring 
image. 
In a resolution optimized analyzer, large energy dispersion and small aberration are 
the main targets. The source size can be limited by an entrance aperture if the source
intensity is sufficient. Otherwise, the finite size effect must be seriously considered.
The spherical aberration can be investigated by minimizing $%
\Delta L$ with respect to the input angular spread $\Delta \theta$.  To 
first order, i.e. the first order focusing, the coefficient of the $\mu =1$ 
term in Eq.(\ref{eq:Taylor}) has to be zero. Second 
order focusing requires $\mu=1$ and $\mu=2$ terms to be zero. 
Principally three free parameters allow us to achieve third order focusing, but in practice
usually only first order focusing is available due to the fact that the focusing behavior depends
strongly only on the voltage configuration of the analyzer. 
In a transmission optimized analyzer, 
large acceptance ($\Omega $) and source size ($l$) are required so that more particles
 can be accepted by the analyzer. When the spherical aberration is controlled such that the 
size of the aberration is smaller than the image of the finite source, demagnification is 
favored to keep high transmission. 

Lacking graphical illustration, pure trajectory analysis does not
volunteer all the information for optimization. For instance, it has been pointed out by
Hafner et al. \cite{Hafner68} that the minimum beam-width does not always occur at the image in the
case of second order focusing, but that the beam can be narrower before it comes to
focus at the image. More importantly, practical applications involve the fringe
fields which often can not be modeled properly, and even when they
do, the analytical equations become too complicated for intuitive
interpretation.

With the advent of modern computers, Poisson equation can be solved numerically using
finite difference or finite element methods. 
The trajectories of charged particles can be calculated very accurately 
 assuming that the boundaries
of the fields are properly setup, the integrating steps are reasonably fine,
and convergence tests are performed during the integration. Thus, it is
possible to include the fringe fields and their use to minimize the aberration by tracing
the minimum beam-width numerically. New, superior modes of
operation can be found through this method and by adding entrance and
exit lenses, the image quality can be fine-tuned\cite{Ross94}. The
`minimum-width' ray tracing is often performed either along the symmetric axis or
its perpendicular plane to study the aberration.
As far as we know, most published results focused their attentions
on the spherical aberration behavior only. 
The electron optical behavior of analyzers has several distinct features which are not seen 
in the performance of rotationally symmetric systems, as can be seen in the
 development of electron microscope
\cite{sedlacek96}.
Due to the fact that the principle trajectory is curved in the analyzer,
the up-down symmetry is broken. This has several significant consequences. 
For example, in rotationally symmetric system, the second-order spherical aberration is
 eliminated by symmetry; while in curved system all orders of spherical aberration can be present.
The astigmatism in rotationally symmetric system is only a second-order effect; in curved
 system it is intrinsic. The spherical aberrations in curved system can easily 
be very significant, even overpowering the finite size effect, unless special efforts are 
made to suppress them. It was found by Ross et al.\cite{Ross94} that the aberration behavior
 can be controlled by adding Herzog lenses to modify the fringe fields of the analyzer.
 We found that even when the source is moved significantly along the principle trajectory, the aberration of the SDA remains in the same mode. Thus we are able to regard the spherical aberration as a minor effect once 
second-order focusing is achieved, and advance our prescription to include the 
magnification of the finite source. As will be shown later, the 
image plane in the curved system may not be perpendicular to the principle trajectory. 
The magnification factor will be more properly presented by a matrix rather than a scalar. 
Therefore looking in only one direction in the ray-tracing results will not provide 
the full information of the imaging process. This is even more relevant when 
an asymmetric field is applied along the flight path. The rotation of the image
by the asymmetric fields can improve the resolution compared to
the symmetrical configuration. 

\section{Imaging Matrix Approach}
\subsection{Formula and Evaluations}
We extend the scalar field representation of Eqs.(\ref{eq:dispersion}) and (\ref{eq:resoln1}) to 
a vector field representation as follows.

If $X$ is the coordinate representing the image,

$x$ the coordinate representing the object,

$\alpha$ the half angle of the exit beam with energy $E$,

$l$ the source finite size,

$M$ the magnification factor, 

$M_l$ the lateral magnification, 

$\widehat{T}$ the direction of the chromatic image,

$\widehat{L}$ the direction of the image,

and $\widehat{n}$ unit vector normal to beam trajectory,

\noindent
then the defocusing broadening due to the rotation of images is given by $\sqrt{M^{2}-M_l^{2}} 
l\alpha $,

the dispersive field by $\overrightarrow{\nabla }E=\frac{dE}{dL}\widehat{T}$,

the chromatic aberration by $\overrightarrow{\nabla }E\cdot \widehat{n}$, and finally 

the defocusing broadening due to chromatic aberration by $\left( \frac{\Delta E}{%
\overrightarrow{\nabla }E}\cdot \widehat{T}+\frac{1}{2}M\widehat{L}\cdot 
\widehat{T}\right)\cdot\alpha $.

These relations are depicted in Figure \ref{fg:orient}.
In general, the imaging process can be expressed as
\begin{eqnarray}
X_{i}=m_{ij}x_{j}+\varepsilon _{ijk}x_{j}x_{k}+\gamma
_{ijkl}x_{j}x_{k}x_{l}+... \label{eq:img}
\end{eqnarray}
To the first order, the image is constructed
by linear mapping of the object. Thus $m_{ij}$ can be represented by a matrix
which characterizes the imaging properties. To higher order, non-vanishing $%
\varepsilon _{ijk}$ , $\gamma _{ijkl}$ ... terms carry the information on 
geometric aberrations. Although the formulation does not include the
spherical (angular) aberration, this effect can be easily taken into account
as a universal blurring in the image. Aberration is included in first order 
imaging as 
\begin{eqnarray}
X_{i}=m_{ij}x_{j}\pm \Delta _{i}\left( \alpha \right) \label{eq:aberr}
\end{eqnarray}
where $\Delta $ depends only on the acceptance half angle 
($\alpha$) of the instrument. In our case, $\alpha =3.5^{\circ }$ and $l=0.5mm$, less
than 5\% contribution to the minimum beam width is based on the  aberration after the 
second order focusing is achieved. To
determine the matrix elements, point sources positioned along several $x$
positions are set up as the input rays for the instrument. After the rays
within an emission angle $\tilde{\alpha}$ traverse through the analyzing field, 
each bundle of exit rays are traced to
find the minimum `beam-width', establishing the image position. 
It is worth mentioning that in preparing the input condition, $\tilde{\alpha}$
should not be too big, for the aberration may effect the result. We choose to
take the semi-angle $\tilde{\alpha}=$0.65$^{\circ }$, 15 rays,  and $\Delta r\left( \Delta z\right) =1mm$ for
all simulation in this investigation.
In our SDA-90, the incident angle is 48$^{\circ
}$, and the exit angle is 42$^{\circ }$. (Fig. \ref{fg:sda90}). Following the convention used by
Ross et al., the source coordinate is defined as $d\vec{l}=d\vec{r}+d\vec{z%
}$, and the image coordinate $d\vec{L}=d\vec{R}+d\vec{Z}$ respectively. The imaging matrix $\textbf{m}$ relates the source vector to the image vector as 
\begin{eqnarray}
\left( 
\begin{array}{l}
\Delta R \\ 
\Delta Z
\end{array}
\right) =\left( 
\begin{array}{ll}
m_{Rr} & m_{Rz} \\ 
m_{Zr} & m_{Zz}
\end{array}
\right) \left( 
\begin{array}{l}
\Delta r \\ 
\Delta z
\end{array}
\right) \label{eq:matrix}
\end{eqnarray}
The magnification factor $M$ can be calculated as ${\sqrt{\Delta R^{2}+\Delta Z^{2}}}
/ {\sqrt{\Delta r^{2}+\Delta z^{2}}}$.
For a unit vector perpendicular to
the incident beam trajectory(principle ray), the lateral magnification
applied to our SDA is, according to Figure \ref{fg:orient},
\begin{eqnarray}
M_l=\left( \sin 48^{\circ },\cos 48^{\circ }\right) \cdot \left( 
\begin{array}{ll}
m_{Rr} & m_{Rz} \\ 
m_{Zr} & m_{Zz}
\end{array}
\right) \left( 
\begin{array}{l}
\sin 42^{\circ } \\ 
-\cos 42^{\circ }
\end{array}
\right) \label{eq:matrix2}
\end{eqnarray}
To calculate the resolution, the dispersive curve at the proximity of the
image point has to be calculated. This can be achieved easily by carrying 
out the ray tracing for a
bundle of rays with the same input conditions but different energies. The
energy dispersion due to the analyzing field can be calculated by taking 
cuts on the ray bundle along $\widehat{n}$ direction at the proximity of 
the image. When $\alpha$ is large, 
including the defocusing effect, the chromatic
images of different energy bundle emerging from the same source point must be 
traced. Again all the chromatic images line up linearly based on the first order approximation. 
Figure \ref{fg:orient} also shows typical inclination ($\widehat{T}$) of the chromatic
image. The resolution of the analyzer can be estimated following the convention:
size of chromatic aberration=size of the image 
\begin{eqnarray}
\frac{\Delta E}{\overrightarrow{\nabla }E}\cdot \widehat{n}+\left( \frac{%
\Delta E}{\overrightarrow{\nabla }E}\cdot \widehat{T}+\frac{1}{2}M\widehat{L}%
\cdot \widehat{T}\right) \cdot \alpha =\sqrt{M^{2}-M_l^{2}}l\alpha +M_l l +\Delta_{spher.aberr.}(\alpha) \label{eq:resoln2}
\end{eqnarray}
When $\alpha$ is small, neglect the defocusing and spherical aberration, $\Delta
E=M_l l\times \left( \overrightarrow{\nabla }E\cdot \widehat{n}\right) $. The
advantage of the analyzer under investigation is the minimization of the right-hand
side due to rotation of images achieved with a non-symmetric analyzing
field.

\subsection{Results and Discussions}

Table I shows the results obtained at three source positions that are relevant to 
us by the techniques outlined above for the symmetric
and asymmetric modes. The voltage settings are optimized according to the aberration curve
such that second order focusing is obtained in both modes.
Note that while the overall maginifaction factors ($M$) in the asymmetric mode may be larger, 
the lateral magnification factors ($M_l$) are always smaller than those in the symmetric 
mode. A series of evaluations was done at source positions from $s=-180mm$ to $s=80mm$. 
Figure \ref{fig:vector} shows the 
graphical representation of the results. Rotation of image happens in both cases. 
In the symmetric mode, however, the rotation (presented by the dashed lines) is 
pretty much confined to the same 
quadrant, although an inversion is caused by the fringe fields when the source 
is placed so close that higher order effects contribute significantly. 
Figure \ref{fig:all4} details the imaging properties for both operational modes. We found 
that the asymmetric mode generally provides a smaller $M_l$ (remember this is the projection of the image perpendicular to the principle trajectory) thanks to the rotation of image. 
That is to say, the charged particle flux will
be better focused seen by a slit mounted perpendicularly to the principle trajectory, 
leading to higher transmission. Due 
to the fact that part of the image will not lie in the slit plane, some defocusing
broadening will happen. The defocusing effect involves the extension of the source image
 along the principle trajectory and the angular spread of the flux. Both factors in the 
 asymmetric mode enhance the defocusing broadening of the outgoing flux. However since the 
size of the image is so small, for reasonable
angular dispersion, the defocusing broadening is insignificant.

The dramatic changes for $s > 50 mm$ in the magnification factor and orientation of the images
( Figs. \ref{fig:vector}, \ref{fig:all4}(b)(d)) are due to the presence of fringe fields. 
Particularly in the symmetric case, it creates long tails in the transmission function which are 
not easy to control.  
Figure \ref{fig:resoln} depicts the resolution  $D^{-1} =M_{l}({\nabla }E\cdot \widehat{n})$
 as a function of the source position for both modes. The resolution of 
the analyzer is the same for
 both cases when the source is exactly on the symmetric axis($s=0$). For $s > 0$ 
the resolution of the asymmetric mode is better; while for $s < 0$ the symmetric 
mode is preferred. In the UTA neutrino unit, the source is a cylindrical cell, 32 mm wide, 
35.5 mm high with a ring opening of 0.5mm. Most of the beta flux will come from 
the $s=0mm$ to $s=21.5mm$ sector in the cell. Operating the analyzer 
in asymmetric mode will improve the transmission by $\sim 23 \%$ and resolution
 by $\sim 4\%$ at the stage as an preanalyzer. Since the resolution of the CMA 
depends linearly on the size of lateral image formed by SDA, the overall 
resolution of the system gained by operating the SDA in asymmetric mode is over $50\%$.  
In the course of this investigation, changing 
the source position did not alter the second order focusing in either case. 
When the source is positioned in the field free region, the position of the image is 
largely decided by the analyzing fields provided by SDA. Tests were carried out to check the
influence of biasing voltages of the SDA and the Herzog lenses on the 
position of the image for both 
symmetric and antisymmetric cases. Same amount of voltage is added or subtracted from SDA 
and lenses up to hundreds of volts. The image shift due to modified Herzog lenses is an order 
of magnitude smaller than equivalently modified SDA fields. However, the fringe fields between the 
lenses and SDA will change the aberration curve completely, and participate strongly
in the rotation of images. Fundamentally, the rotation during the imaging is an
inherent property for any curved optical system, and its treatment can still reside 
in the general paraxial principle with special care for its vectorial nature.  

\section{Experiments}

A Steigerwald type gun \cite{Steigerwald} is chosen as an electron beam source because
the monochromaticity of the beam is better than
0.01\% \cite{Zhang} and an adjustable real image, created beyond the electron gun  
by telefocusing, can be used as the input object
for the SDA. The electron source is used to measure the action of the dispersive field of 
the SDA, and thus makes the verification of
the calculated imaging property of the SDA possible. While the position of the object
moves--by adjusting the position of the inner Wehnelt cylinder of the
gun--the image is traced both vertically and
horizontally. Since the object will be in the
proximity of the symmetric axis, the SDA will form a slightly magnified lateral image in the vertical direction.

\subsection{Setup}

A self-biasing electron gun based on Steigerwald's design is mounted on a
rotatable frame under the SDA-90. The incidence angle of the gun can be
adjusted. A Faraday cage mounted at the entrance of SDA, 16mm offset from
symmetrical axis, with 50 micron aperture facing the incident e-beam acts
as a beam `checker'. Passing the SDA-90, the electrons face a similar Faraday cage(detector)
 which is mounted on another rotatable frame with moving mechanisms which allow the
cage to move in the plane perpendicular to the exit beam. The two rotating frames
are aligned horizontally by a digital indicator to better than 1 arc minute.
In the vertical direction two axes are aligned optically through a
laser beam defined by two apertures and a photodiode with a 100 micron
entrance aperture. The accuracy is within 50 micron. The setup is detailed in Figure
 \ref{fg:instrum}. The incidence angle of the gun is first
fixed by the `checker'. Then the gun is rotated 90$^{\circ }$ to allow the
beam to go through the SDA. By adjusting the position of the inner Wehnelt
cone of the electron gun, beams with different sizes and points of
convergence can be created. The beam profiles are first measured with the beam
checker before entering the SDA, and later the exit beam profiles are
measured again. The checker only measures the horizontal
beam-width; while the detector measures both horizontal and vertical
profiles. The cage current is measured by a Ketheley 616 multi-meter
followed by a Dell P100 computer through GPIBs. The power supply
of the electron gun is Spellman RHSR60N. A Bertan 205B power supply sets the 
voltage of the inner sphere of SDA, and a Fluke 408B power supply for Herzogs lenses.
 The beam intensities range from 2 to
6 micro-amps. The whole chamber is maintained at $\sim $2$\times 10^{-6}$
Torr. The vacuum tank is shielded by mu-metal and the transverse magnetic fields are
 measured to be in the range of mGauss.
\subsection{Results}
The principle trajectory is determined by the incidence angle of the electron beam,
the checker's z position, the inclination angle of the detector, and the detector Z position. The 
voltages of the SDA and Herzog lenses are set according to the simulation. 
The variables are the azimuthal position of the detector and the incident energy of the electron.
Although the electron energy can be recored to 6 digit precision by a differential voltmeter, 
its absolute value can only be read out in three significant digits. The agreement between the
relativistic numerical calculation and the experimental results of the electron energy is within $0.5\%$.
We also found that the principle trajectory is rather insensitive to the voltage setting of 
the two Herzog lenses. The beam envelopes under investigation all have excellent gaussian shape(Fig. \ref{fg:gauss}). No observable baseline fluctuation appears in the measured beam intensity.
The energy resolution of the analyzer is measured by adjusting the incident
beam energy from 19940 eV to 20060 eV, such that the gaussian beam can scan through the detecting aperture. Depicted in Figure \ref{fg:eng_res}, the full width at half maximum(FWHM) of the beam profile defines the energy resolution to be $\Delta E/E =2.47 \times 10^{-3}$. 
This is measured at incident beam size of 634 micron
and exit beam size of 726 micron. The beam spread is only 2.5 mrad; thus no angular
aberration is seen. Since $\Delta E/E=M_{l} ({\nabla }E\cdot \widehat{n})/E$, inferred by this result, ${\nabla }E\cdot \widehat{n}= 68eV/mm$. This measurement agrees with the ray
trace result ($70eV/mm$) very well.
The focusing property of the SDA is elucidated in terms of the ratios of the beam-widths measured before and after the SDA. Depicted in Figures \ref{fg:sda21} and \ref{fg:sda22}, both vertical and horizontal lateral beam-widths were measured, at $z=87.35$ by the checker and at $Z=522.52mm$ by the detector.  Since the ray-tracing predicted several optical properties of 
the SDA very well, we use the beam-width ratio to calculate the position of the object. 
Particularly we examine the data obtained by setting inner Wehnelt position marked $0$. In Figure \ref{fg:sda21}, vertically the detector measures 726 micron lateral
beam-width (FWHM) while the input beam-width is 634 micron (beam spread
3.12 mrad, point of convergence is 203mm ahead of the checker, based on the previous measurements on the focusing property of telefocus gun. ) at the checker.
Applying the simulation results in Figure \ref{fig:all4} and  taking into account the modification due to the defocusing, the object is found at $z=61.2 mm$, and 
correspondingly the image is at $Z=483mm$. ( Complete results see the previous 
article.) This result will on one hand be used in the
 previous article as another supporting evidence for the proper analysis of the 
emission optics of the 
Steigerwald type electron gun; on the other hand the focusing property of the 
horizontal direction, which is not calculated in the 2D ray tracing, can be 
constructed. From the same input beam, the detector receives a horizontal beam-width
 of 804 micron(FWHM)( Figure \ref{fg:sda22}). Based on the conservation of 
brightness, the beam dispersion angle in this direction is estimated to be 3.49 
mrad. This puts an upper limit 1.27 on the lateral magnification factor in the 
horizontal direction.  Thus separate images will form vertically and horizontally
, featuring the astigmatism of the SDA. However, since eventually a ring source will be used 
in the experiment, we have to look at the vertical image only. The ratio of beam-widths 
measured in both directions is close to 1; this suggests that the broadening in the 
ring image due to the horizontal defocusing will not be significant in our case. 
This astigmatism is often unchecked by two dimensional ray tracing for cylindrically 
symmetric analyzers and thus could cause anomalous broadening in practice\cite{Kittel98}.
\section{Conclusion}
In this work, we have used the imaging matrix as a tool to evaluate the resolution 
and transmission characteristics of a spherical deflecting analyzer. Results are shown for 
both symmetrically and asymmetrically charged cases. The asymmetrical case is found
to be superior. A SDA-90 has been built based on the simulations. The principle trajectory and the 
dispersion field are checked by experiment using a telefocus gun as the beam source.
 Many details made possible by the imaging matrices provides a straightforward 
database for convoluting the spectrometer function into the finite source. Although 
extensive literature is available on this topic, their results contribute mainly
to cases where the source is small. Thus the aberration behavior is the key 
element to optimize the resolution. Typically a third order aberration curve from analyzer
 of our size ($\sim 50cm$ ) has second order focus of 20 micron. For any finite source which 
is larger than 50 micron, the finite size effect often dominates, and the optimization
 based on the aberration only may not be correct. The imaging matrix approach proposed 
here provide a way of dealing with such problems.

\section{Acknowledgement}
The authors wish to express their gratitude to L.H. Thuesen, H.F. Wellenstein for their 
involvement in the early stage of this work. Special thanks go to the UTA Physics machinists
for their excellent works. This work was supported by Texas Advanced 
Research Project and Robert A. Welch Foundation.

\newpage
%\listoffigures

\begin{table}

\begin{tabular}{||l||l|l||}
\hline
& Asymmetric & Symmetric \\
\hline
\hline 
Source Position & 
\begin{tabular}{l}
$V_{out}=0V$ \\ $V_{in}=15700V$ \\ $V_{H}=4500V$ 
\end{tabular}
&
\begin{tabular}{l} 
$V_{out}=8922V$ \\ $V_{in}=6588V$ \\ $V_{H}=-6580V$ 
\end{tabular} 
\\
\hline
(1) r=0 & 
\begin{tabular}{l}
$\mathbf{m}^{\left( 1\right) }=\left( 
\begin{array}{ll}
-0.84 & -0.84 \\ 
-0.39 & 2.4
\end{array}
\right) $ \\ $M_{l}^{\left( 1\right) }=-1.32$\\ $M^{\left( 1\right) }=2.0$ 
\end{tabular}
&
\begin{tabular}{l}
$ \mathbf{m}^{\left( 1\right) }=\left( 
\begin{array}{ll}
-1.076 & -0.900 \\ 
0.284 & 2.636
\end{array}
\right) $\\ $M_{l}^{\left( 1\right) }=-1.476$\\ $M^{\left( 1\right) }=2.150$
\end{tabular}
\\
\hline 
(2) r=16 & 
\begin{tabular}{l}
$\mathbf{m}^{\left( 2\right) }=\left( 
\begin{array}{ll}
-0.637 & -1.83 \\ 
-0.86 & 3.76
\end{array}
\right) $ \\ $M_{l}^{\left( 2\right) }=-1.56$ \\ $M^{\left( 2\right) }=3.5$ 
\end{tabular}
&
\begin{tabular}{l} 
$\mathbf{m}^{(2)}=\left( 
\begin{array}{ll}
-1.918 & -1.390 \\ 
0.224 & 3.648
\end{array}
\right) $ \\ $M_{l}^{\left( 2\right) }=-1.900$ \\ $M^{\left( 2\right) }=2.573$
\end{tabular}
\\ 
\hline
(3) r=22.3 & 
\begin{tabular}{l}
$\mathbf{m}^{\left( 3\right) }=\left( 
\begin{array}{ll}
-0.50 & -2.40 \\ 
-1.15 & 4.49
\end{array}
\right) $ \\ $M_{l}^{\left( 3\right) }=-1.67$ \\ $M^{\left( 3\right) }=4.36$ 
\end{tabular}
&
\begin{tabular}{l} 
$\mathbf{m}^{\left( 3\right) }=\left( 
\begin{array}{ll}
-2.425 & -2.172 \\ 
0.552 & 4.784
\end{array}
\right) $ \\ $M_{l}^{\left( 3\right) }=-2.138$\\ $M^{\left( 3\right) }=3.186$
\end{tabular}
\\
\hline
\end{tabular}
\caption{The imaging matrix for three source positions ($r$ in mm). 
Results are obtained for both symmetric and asymmetric configurations. }
\end{table}

%\newpage
\vspace*{1cm}
\begin{figure}[htbp]
\caption{The SDA-90 is ray-traced from an emitting point O at location (z,r) in a cylindrically symmetric configuration. The principle trajectory has incident angle $48^\circ$ and exit angle $42^\circ$. $s$ is defined as the distance from O to symmetric axis along the principle trajectory. The image, usually located in $R<0$ plane,  can be ray-traced by extrapolating the exit rays.}
\label{fg:sda90}
\end{figure}

%\newpage
\vspace*{1cm}
\begin{figure}[htbp]
 \begin{center}
\caption{This graph shows the typical orientation of the magnified image ($\widehat L$)
and chromatic image ($\widehat T$). }
\label{fg:orient}
 \end{center}
\end{figure}

%\newpage
\begin{figure}[htbp]
\begin{center}
\caption{The magnification vectors oriented in R-Z plane. The unit circle and
 the principle exit trajectory are also shown in the graph. The arrows represent the image vectors corresponding to (a) s=-100mm, (b) s=-30mm, (c) s=0mm, (d) s=25mm,
 (e) s=60mm, (f) s=75mm.}
\label{fig:vector}
\end{center}
\end{figure}

%\newpage
\begin{figure}[t]
\begin{center}
  \caption{(a)The image position, Z, (b)lateral magnification, $M_l$, (c)angular magnification, $M_\alpha$ (d)the image orientation relative to the $\widehat{Z}$ axis, $\Theta$, calculated by varying the source position $s$ from $-180$ to 80 mm. The framed area corresponds to the tritium gas cell to be installed in UTA neutrino unit}
\label{fig:all4}
\end{center}
\end{figure}

%\newpage
\begin{figure}[htbp]
  \begin{center}
    \caption{The resolution curve, $D^{-1} =M_{l}({\nabla }E\cdot \widehat{n})$. Note the defocussing and spherical aberration are neglected.}
    \label{fig:resoln}
  \end{center}
\end{figure}

%\newpage
\begin{figure}[htbp]
\caption{Experimental apparatus,(A) Rotatable Steigerwald type electron gun 
(B) Faraday cage and aperture (checker)(C) movable Faraday cage and apertures 
(detector) (D) SDA (E) digital indicator (F)laser (G) centering photodiode (H) Mu-metal (I) optical encoder.}
\label{fg:instrum}
\end{figure}

%\newpage
\begin{figure}[htbp]
  \begin{center}
    \label{fg:gauss}
    \caption{Typical beam scan from the detector.}    
  \end{center}
\end{figure}

%\newpage
\begin{figure}[htbp]
  \begin{center}
    \caption{The resolution curve measured by 726 micron electron beam.
 The error bar is magnified by a factor of 10 to show up in the scale.}
    \label{fg:eng_res}
  \end{center}
\end{figure}

%\newpage
\begin{figure}[htbp]
  \begin{center}
\caption{The beam FWHM measured in the vertical direction. Each step of inner Wehnelt position is 0.27mm.}
\label{fg:sda21}
\end{center}
\end{figure}

%\newpage
\begin{figure}[htbp]
  \begin{center}
\caption{The beam FWHM measured in the horizontal direction.}
\label{fg:sda22}
\end{center}
\end{figure}


\newpage
\begin{thebibliography}{99}
\bibitem{Lobashev94}A.I.Belesev et al., Phys. Lett. B \textbf{350}, 263 (1995)
\bibitem{Bach93}H. Bache, et al., Nucl. Phys. B (Proc. Suppl.) \textbf{31}, 46 (1993)
\bibitem{Decman91}D.J. Decman and W. Stoeffl, Phys. Rev. Letters \textbf{64}, 2767 (1991)
\bibitem{Robertson91}R.G.H. Robertson, et al., Phys. Rev. Letters \textbf{67}, 957 (1991)
\bibitem{Wilkerson93}J.F. Wilkerson, Nucl. Phys. B (Proc.Suppl.) \textbf{31},32 (1993)
\bibitem{Aston19}F.W.Aston, Philos. Mag. \textbf{38}, 710(1919)
\bibitem{Purcell38}E.M. Purcell, Phys.Rev. \textbf{54},818(1938)
\bibitem{Ashby58}N. Ashby, Nucl.Instrum. \textbf{3},90(1958)
\bibitem{Kessler64}J. Kessler, and N. Weichert, Nuclear Instruments and Methods \textbf{29}(1964)
\bibitem{Ross94} A.W.Ross, L.K. Smith, Changde Xie, L.H. Thuesen, M. Fink and H.F. Wellenstein,J. Elec. Spectr. and Rel. Phenom. \textbf{69},189(1994)
\bibitem{Hafner68}H. Hafner, J.A. Simpson, and C. E. Kuyatt, Rev. Sci. Instrum. \textbf{39}, 33(1968)
\bibitem{sedlacek96} M. Sedlacek, Electron Physics of Vacuum and Gaseous Device, John Wiley \& Sons, INC, 1996.
\bibitem{Steigerwald} K.H. Steigerwald, Optik \textbf{5}, 469(1949)
\bibitem{Zhang} J.D. Coffman, M. Fink and H. Wellenstein, Phys. Rev. Lett. \textbf{55}, 1392(1985)
\bibitem{Kittel98} M.E. Kittel, Master Thesis, UT-Austin 1998 
\end{thebibliography}
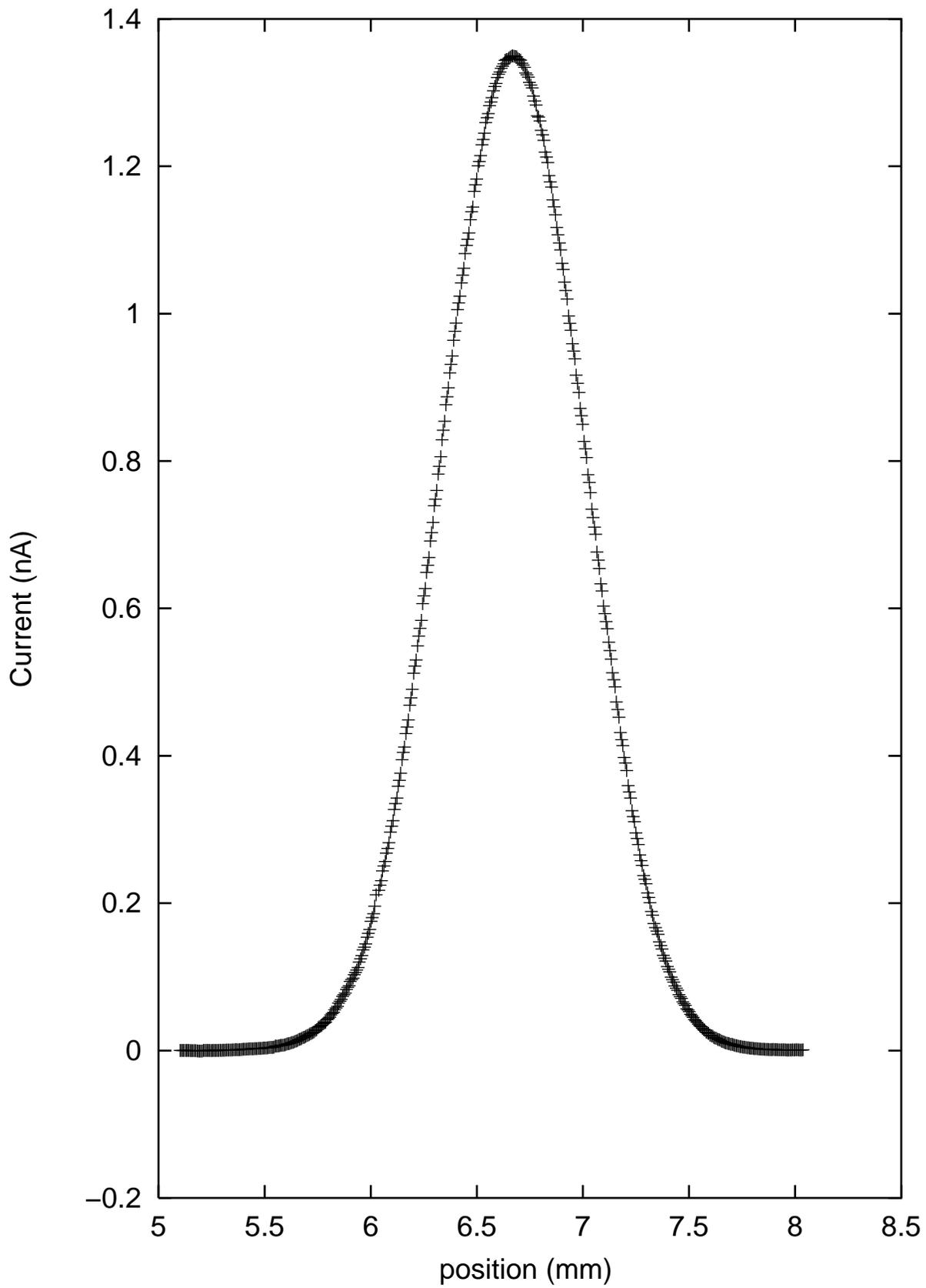
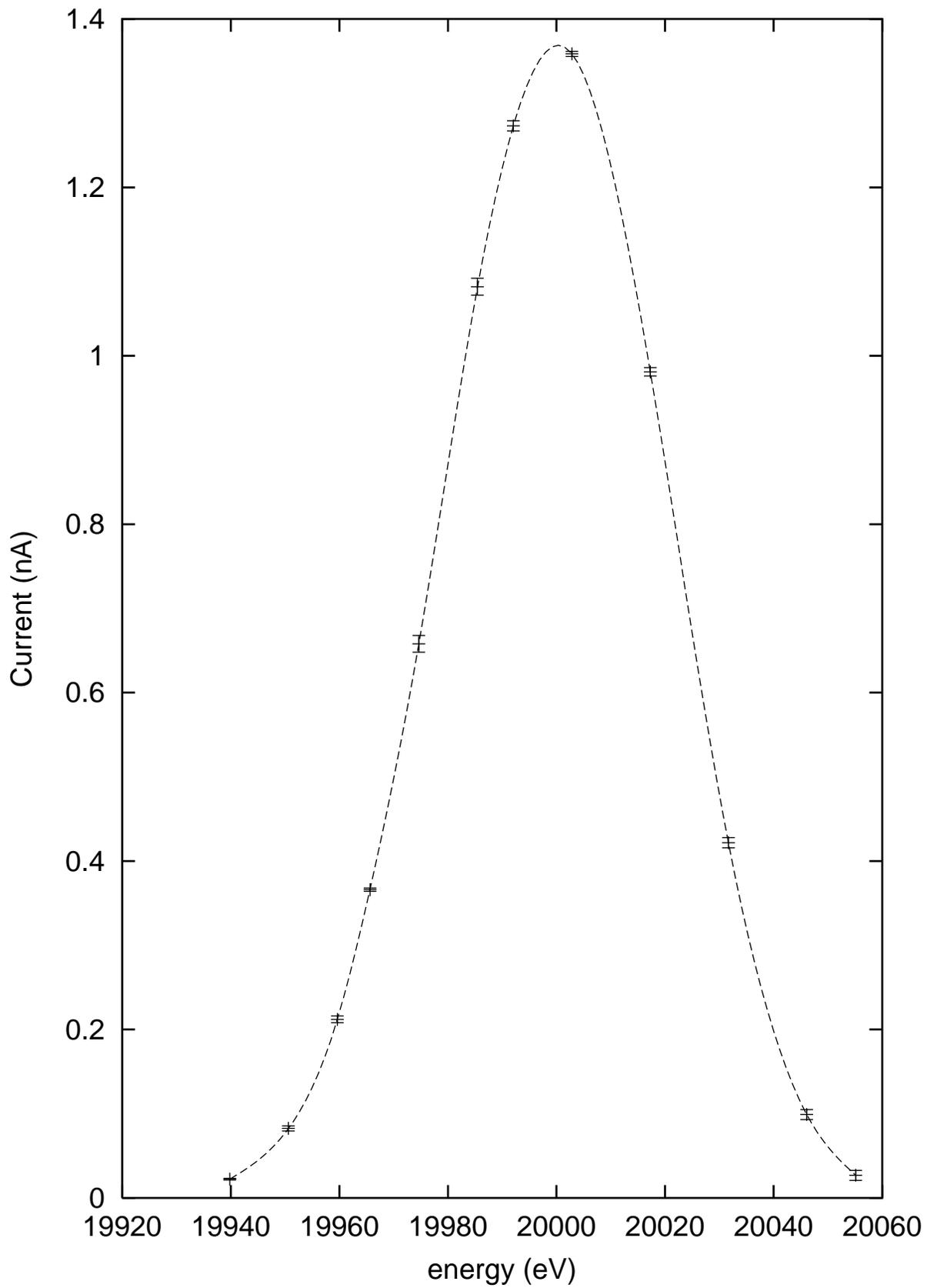
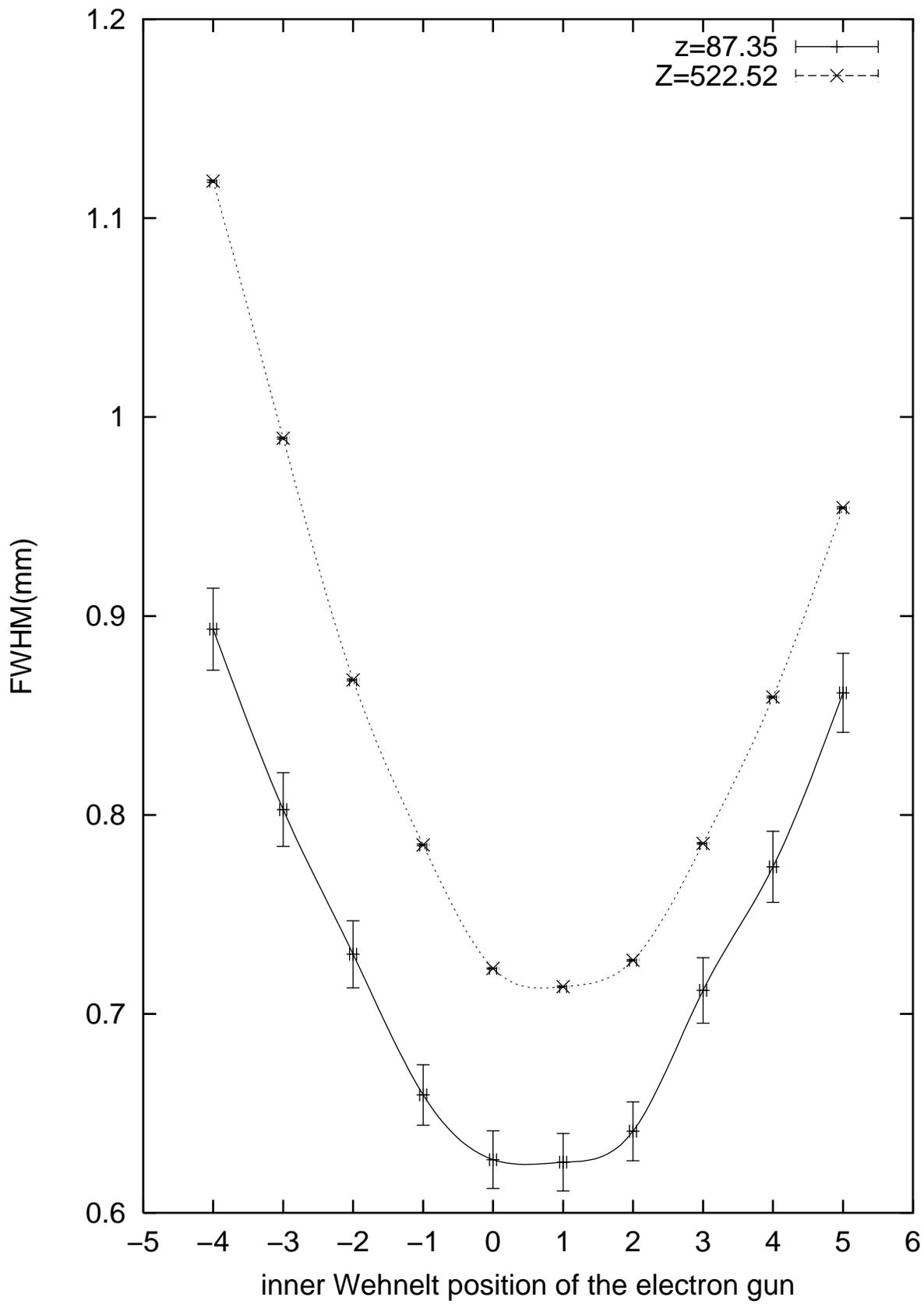
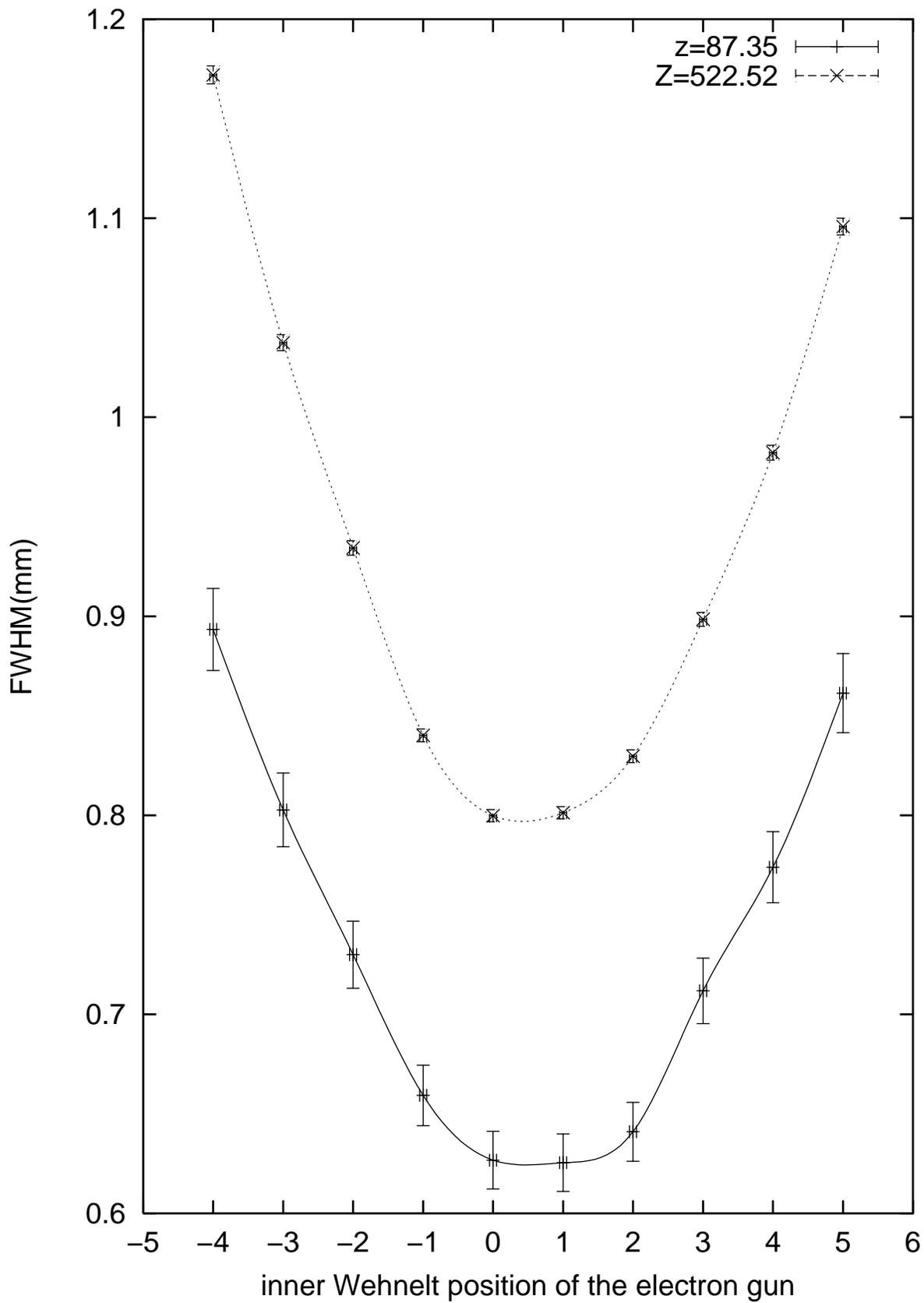
\end{document}